\documentclass[apjl, revtex4]{emulateapj}
\usepackage{pdflscape}

\usepackage[backref,breaklinks,colorlinks,citecolor=blue]{hyperref}
\usepackage[all]{hypcap}

\begin{document}

\submitted{Submitted to ApJL}

\shorttitle{Confirming the Quiescent Galaxy Population out to $z=3$}
\shortauthors{Man et al.}

\title{
Confirming the Quiescent Galaxy Population out to \lowercase{$z=3$}: \\
A Stacking Analysis of Mid-, Far-Infrared and Radio Data}

\author{Allison W. S. Man\altaffilmark{1,2},
Thomas R. Greve\altaffilmark{3}, Sune Toft\altaffilmark{1}, 
Benjamin Magnelli\altaffilmark{4}, Alexander Karim\altaffilmark{4},
Olivier Ilbert\altaffilmark{5}, Mara Salvato\altaffilmark{6},
Emeric Le Floc'h\altaffilmark{7}, Frank Bertoldi\altaffilmark{4},
Caitlin M. Casey\altaffilmark{8}, Nicholas Lee\altaffilmark{2}, Yanxia Li\altaffilmark{2}, 
Felipe Navarrete\altaffilmark{4},
Kartik Sheth\altaffilmark{9}, Vernesa Smol\v ci\'c\altaffilmark{10}, 
David B. Sanders\altaffilmark{2}, Eva Schinnerer\altaffilmark{11}, and Andrew W. Zirm\altaffilmark{1}}
\email{allison@dark-cosmology.dk}

\altaffiltext{1}{Dark Cosmology Centre, Niels Bohr Institute, University of Copenhagen, Denmark}
\altaffiltext{2}{Institute for Astronomy, 2680 Woodlawn Drive, University of Hawaii, Honolulu, HI 96822, USA}
\altaffiltext{3}{Department of Physics and Astronomy, University College London, Gower Street, London WC1E 6BT, UK}
\altaffiltext{4}{Argelander-Institut f\"ur Astronomie, Universit\"at Bonn, Auf dem H\"ugel 71, D-53121 Bonn, Germany}
\altaffiltext{5}{Aix Marseille Universit\'e, CNRS, Laboratoire d'Astrophysique de Marseille, UMR 7326, F-13388 Marseille, France}
\altaffiltext{6}{Max-Planck-Institut f\"ur extraterrestrische Physik, Garching bei M\"unchen, D-85741 Garching bei M\"unchen, Germany}
\altaffiltext{7}{Laboratoire AIM, CEA/DSM/IRFU, CNRS, Universit\'e Paris-Diderot, 91190 Gif, France}
\altaffiltext{8}{Department of Physics and Astronomy, University of California, Irvine, CA 92697, USA}
\altaffiltext{9}{National Radio Astronomy Observatory, 520 Edgemont Road, Charlottesville, VA 22903, USA}
\altaffiltext{10}{Physics Department, University of Zagreb, Bijeni\v cka cesta 32, 10002 Zagreb, Croatia}
\altaffiltext{11}{Max-Planck-Institut f\"ur Astronomie, K\"onigstuhl 17, D-69117 Heidelberg, Germany}
\altaffiltext{$\dagger$}{Herschel is an ESA space observatory with science instruments provided by European-led Principal Investigator consortia and with important participation from NASA.}

\begin{abstract}
We present stringent constraints on the average mid-, far-infrared and radio emissions
of $\sim$14200 quiescent galaxies (QGs), identified out to $z=3$ in the COSMOS
field via their rest-frame NUV$-$r and r$-$J colors, and with stellar masses
$M_{\star}=10^{9.8\mbox{--}12.2} \,M_{\odot} $. Stacking in deep \textit{Spitzer} (MIPS
$24\,\mu$m), \textit{Herschel}$^{\dagger}$ (PACS and SPIRE), and VLA (1.4\,GHz) maps
reveals extremely low dust-obscured star formation rates for QGs
(SFR $<0.1\mbox{--}3\,M_{\odot}$yr$^{-1}$ at $z \leqslant 2$ and $<6\mbox{--}18\,M_{\odot}$yr$^{-1}$ at $z > 2$), consistent with the low unobscured SFRs ($<0.01\mbox{--}1.2\,M_{\odot}$yr$^{-1}$) inferred
from modeling their ultraviolet-to-near-infrared photometry. 
The average SFRs of QGs are $>10\times$ below those of star-forming galaxies (SFGs) 
within the $M_{\star}$- and $z$-ranges considered.
The stacked 1.4\,GHz signals (S/N $> 5$) are, if attributed solely to star
formation, in excess of the total (obscured plus unobscured) SFR limits,
suggestive of a widespread presence of low-luminosity active galactic nuclei
(AGN) among QGs. Our results reaffirm the existence of a significant population
QGs out to $z = 3$, thus corroborating the need for powerful quenching
mechanism(s) to terminate star formation in galaxies at earlier epochs.
\end{abstract}


\section{Introduction}
Half of the most massive ($M_{\star} \geqslant 10^{11}\,M_{\odot}$) galaxies at
$z\sim1.5$ have evolved stellar populations and SFRs of only a few $M_{\odot}$
yr$^{-1}$ \citep[e.g.,][and references therein]{Ilbert2013}, suggesting that they have undergone a
rapid build-up of stellar mass followed by an effective phase of
star formation (SF) quenching, probably via AGN feedback
\citep[e.g.,][]{Bower2006, Croton2006}. If significant dust is present in these galaxies,
however, it would imply that the SFRs, inferred from the rest-frame ultraviolet (UV), are
severely underestimated, and that their stellar populations are in fact not
old but simply reddened by the dust. 
Direct far-infrared (FIR) measurements of the dust are therefore essential to
unambiguously assess the level of obscured SF.  A recent \textit{Herschel}
stacking analysis by \citet{Viero2013} found that massive QGs at $z>2$ have
IR luminosities comparable to local ultra-luminous IR galaxies (ULIRGs,
$L_{\rm IR} \geqslant 10^{12}\,L_{\odot}$), inconsistent with the
quiescence inferred from the UV continua \citep[e.g.,][]{Ilbert2013} as well as
their low $24\,{\rm \mu m}$ stacked flux densities \citep{Fumagalli2013, Utomo2014}.
If QGs harbor significant dust-obscured SF,
it would challenge the need for powerful quenching mechanisms.
\begin{figure*}[!tbp]
	\centering
	\includegraphics[angle=0,width=\textwidth]{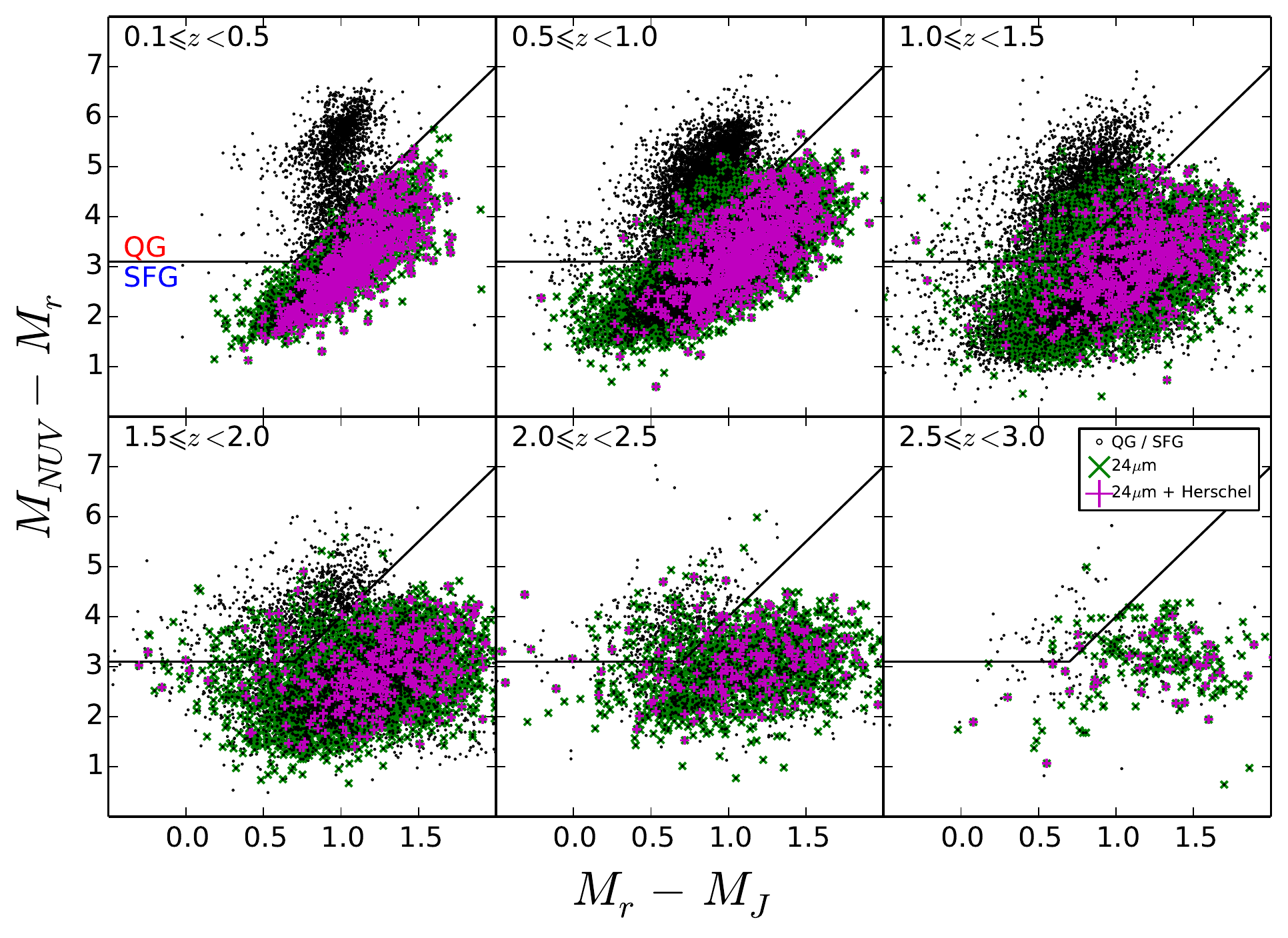}
	\caption{
	Rest-frame NUV$-$r and r$-$J colors for galaxies above the mass-completeness
	limits (small black circles) from the UltraVISTA survey. QGs are defined as having $M_{\rm NUV} - M_{\rm r} > 3 (M_{\rm r} - M_{\rm J})+1$ and $M_{\rm NUV} - M_{\rm r} > 3.1$. 
	The QGs/SFGs classification boundary is marked by black solid lines. Galaxies with
	SFR$_{24}>20\,M_{\odot}$yr$^{-1}$ (green crosses) and \textit{Herschel}
	detections (magenta pluses) are indicated (fractions of the
	total QG sample are listed in Table~\ref{table:detfrac}).
	}
	\label{fig:nuvrj}
\end{figure*}

Here, we analyse a sample of $\sim$14200 QGs with $M_{\star} =
10^{9.8\mbox{--}12.2}\,M_{\odot}$ out to $z=3$, 
selected over 1.48\,deg$^{2}$ in the COSMOS field.
Taking advantage of the available deep multi-wavelength data,
we constrain their dust-obscured SFRs through stacking in \textit{Spitzer}
Multiband Imaging Photometer (MIPS), \textit{Herschel} Photodetector Array
Camera and Spectrometer (PACS; \citealt{Poglitsch2010}) and Spectral and
Photometric Imaging Receiver (SPIRE; \citealt{Griffin2010}) maps. These are
compared with stacks in deep Very Large Array (VLA) radio maps. We infer 
extremely low levels of dust-obscured SF ($<[0.3, 3, 18]\,M_{\odot}$yr$^{-1}$ at
$z\sim[0.8, 1.7, 2.6]$), thus definitively confirming the quiescent nature of these galaxies. 

Magnitudes are quoted in the AB system. We adopt a \citet{Chabrier2003} initial
mass function, and $H_{0}$ = 70\,km\,s$^{-1}$\,Mpc$^{-1}$, $\Omega_\mathrm{M}$ =
0.3 and $\Omega_{\Lambda}$ = 0.7.

\begin{figure*}[!htbp]
	\centering
	\includegraphics[angle=0,width=\textwidth]{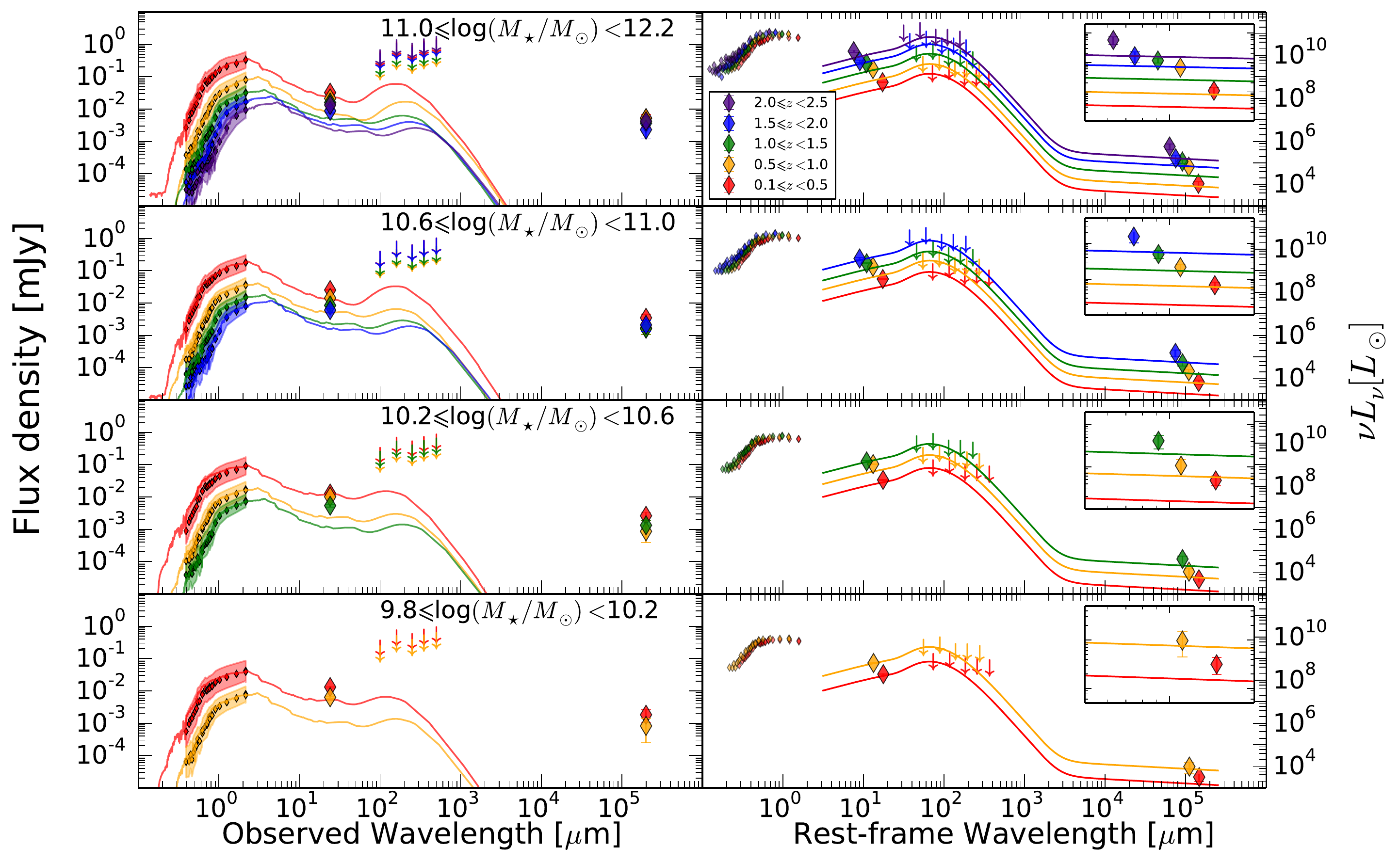}
	\caption{
	Panchromatic SEDs of QGs in four $M_{\star}$- bins (rows) and six $z$- bins
	(colors) in observed (left) and rest-frame (right) frames. The median
	UV-to-near-IR photometry is plotted and shaded with its standard deviations.
	The longer wavelength data represent our stacking results. \textit{Left}:
	At $z>0.5$ the observed $S_{24}$ are higher than that expected from pure
	stellar emissions of elliptical galaxy models of \citealt{Bruzual2003}
	(matched to the median stellar ages from SED fits and scaled to $K$-band
	magnitudes). \textit{Right}: The FIR black-body models \citep{Casey2012}
	fitted to the \textit{Herschel} upper limits (assuming $T_{\rm dust} = 30\,{\rm K}$) are
	co-joined with a radio power-law ($\alpha=-0.8$) and plotted as lines, following
	the radio-FIR correlation presented in \citet{Ivison2010b} with shallow
	redshift evolution.  The templates are not fitted to the radio data.  Shown
	in the insets, the observed $S_{\rm radio}$ is higher than expected from SF.
	The $24\,{\rm \mu m}$ and radio excesses suggest contributions from low-luminosity
	AGN.
	}
	\label{fig:sed}
\end{figure*}

\section{Data and Sample Selection} \label{sec:data}
We select galaxies brighter than ${\rm K_s} = 24$ from the UltraVISTA
survey \citep{McCracken2012} that have $M_{\star} \geqslant 10^{9.8}\,
M_{\odot}$ and photometric redshifts $z_{\rm phot}=0.1-3.0$.
Both $M_{\star}$ and $z_{\rm phot}$ are from \citet{Ilbert2013}, derived from
spectral energy distribution (SED) fits to broadband UV-to-IRAC
photometry \citep{Capak2007,Scoville2007}. A small number of AGN,
identified via their emission in X-rays \citep{Brusa2010, Civano2012}, IRAC bands
\citep{Donley2012}, or the radio \citep{Schinnerer2007, Schinnerer2010}, are removed
to minimise the effects of erroneous SED fits and thus inaccurate
$z_{\rm phot}$ and $M_{\star}$. Including the AGN in the analysis does not change
the stacked flux densities (within the uncertainties) nor the conclusions of
this Letter.

Each galaxy is classified as a QG or a SFG based on its rest-frame NUV$-$r and r$-$J colors (Figure~\ref{fig:nuvrj}).  
NUV$-$r is a measure of the amount of UV light from young stars (i.e.,
recent SF) relative to the red optical light from evolved stellar populations,
while r$-$J constrains the degree of dust attenuation in the red part of the spectrum. 
The QGs are divided into six bins of $z_{\rm phot}$, each of which
is split into four $M_{\star}$-bins (see Table \ref{table:stack}); however, only
$M_{\star}$-bins which are $>90\%$ mass-complete (according to the limits
presented in \citealt{Ilbert2013}) are included.

To weed out dusty galaxies erroneously classified as QGs, 
we cross-correlate our sample with the MIPS $24\,{\rm \mu m}$ catalog of \citet{LeFloch2009} with a 
radius of $2\arcsec$. A redshift-dependent $24\,{\rm \mu m}$ flux density ($S_{24}$) cut-off is
then applied to remove QGs with dust-obscured SFRs $>20\,M_{\odot}$yr$^{-1}$ (as inferred from their $S_{24}$ --- see Section~\ref{sec:ms}).

The fraction ($f_{\rm QG, 24}$) of QGs with $24\,{\rm \mu m}$-inferred SFRs
$>20\,M_{\odot}$yr$^{-1}$ increases with redshift and peaks at
13--19\% for the most massive $z\gtrsim2$ QGs (see Table \ref{table:detfrac}).
This suggests a higher fraction of misclassified QGs at $z\gtrsim2$, which
is unsurprising given their faintness (i $\gtrsim25$).  Overall,
however, the fractions are reassuringly small.
A similar conclusion is reached from the fraction ($f_{\rm QG, H}$) of \textit{Herschel} detected
QGs ($< 6\%$), 
determined using the catalog of \citet{Lee2013} in which the $24\,{\rm \mu m}$ sources are
cross-identified to the \textit{Herschel} detections (i.e., S/N $\geqslant5$ 
in at least two PACS or SPIRE bands).
This population of dusty galaxies having quiescent NUV$-$r and r$-$J
colors could either be SFGs with strong attenuation, or galaxies
containing evolved stellar populations and undergoing rejuvenation of SF
\citep{Lemaux2013}.  The robust \textit{Herschel} detections in the QG region
tend to lie close to the QG/SFG classification boundary at least out to $z=1.5$
(Figure~\ref{fig:nuvrj}), perhaps indicative of their post-starburst nature
\citep{Hayward2014}.

For the stacking analysis (Section~\ref{sec:stacking}) we use the
aforementioned MIPS $24\,{\rm \mu m}$ imaging (FWHM $\simeq6\arcsec$) from
\citet{Sanders2007}, while the \textit{Herschel} PACS and SPIRE maps are from the PACS
Evolutionary Probe survey (PEP; \citealt{Lutz2011}) and the \textit{Herschel}
Multi-tiered Extragalactic Survey (HerMES; \citealt{Oliver2012}), respectively.
The PACS maps reach depths of 5 and 10.3\,mJy\,beam$^{-1}$ ($3\sigma$) at 100 and
160\,$\mu$m, respectively (FWHM $\simeq$ 6.8$\arcsec$ and 11$\arcsec$), and
SPIRE 250, 350, and 500\,$\mu$m depths are 8, 11, and 13\,mJy\,beam$^{-1}$
($3\sigma$), respectively (FWHM $\simeq$ 18.2$\arcsec$, 24.9$\arcsec$, and 36.3$\arcsec$). For the radio stacking we use the 1.4\,GHz VLA-COSMOS large survey
\citep{Schinnerer2007, Schinnerer2010}, which reaches a root-mean-square noise
(rms) of 15\,$\mu$Jy\,beam$^{-1}$ at an angular resolution of $\sim1.5\arcsec$ (FWHM).


\begin{figure*}[t]
	\centering
	\includegraphics[angle=0,width=\textwidth]{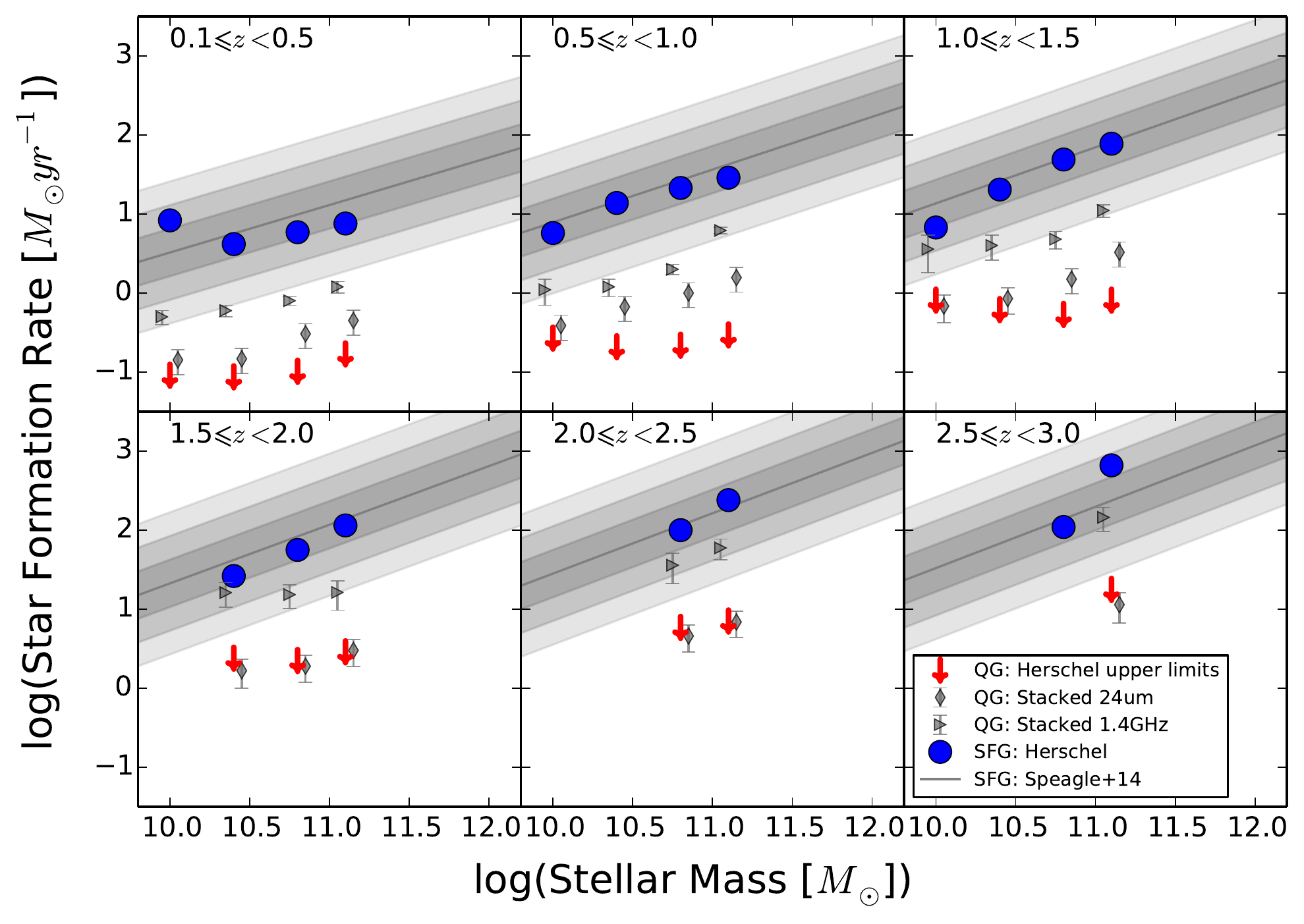}
	\caption{
	 SFRs inferred from stacking as a function of $M_{\star}$ and $z$.  Blue
	 circles and red downward arrows represent the SFRs from global deblending
	 and stacking in \textit{Herschel} for SFGs and QGs respectively, with the
	 latter ones representing $3\sigma$ upper limits for QGs since they are
	 consistent with no detection. Assuming the $24\,{\rm \mu m}$ and radio emissions
	 originate from SF only, we plot the inferred SFRs as gray diamonds and
	 triangles. SFR$_{\rm radio}$, as well as SFR$_{24}$ at $z<1.5$, show clear
	 offsets from SFR$_{\rm H}$ for QGs, therefore part of the radio emission in QGs
	 likely arises from low-luminosity AGN.  The SFR-$M_{\star}$ measured in a
	 recent compilation \citep{Speagle2014} is plotted as gray lines, and the
	 1$-$3 $\times$ observed dispersion ($\sigma_\mathrm{SFR}$=0.3) are shown as
	 dark-to-light shades.
	 QGs have SFRs at least $\gtrsim1$\,dex below those of SFGs out to $z\sim3$.
	}
	\label{fig:ms}
\end{figure*}

\section{Stacking}  \label{sec:stacking}
Our \textit{Herschel} maps are characterised by a high level of source
confusion which, if unaccounted for, will bias a stacked signal \citep{Marsden2009, Bethermin2010, Kurczynski2010, Viero2013}. Here, we use
a global deblending technique similar to that of \citet{Kurczynski2010}
but generalised to deblend multiple galaxy samples simultaneously, which in our
case totaled 87 samples (separated by their SFG/QG classifications, $z$- and
$M_{\star}$-bins, and SFR$_{24}$ threshold).

Source confusion is not an issue for our radio maps due to the high angular
resolution, and the stacked signal of a given sample was determined from the
median combination of the galaxy postage stamps belonging to the sample.  The
MIPS $24\,{\rm \mu m}$ stacks were determined in a similar way, despite the larger beam
size. To ensure that our $24\,{\rm \mu m}$ median stacks were not biased, we stacked
samples of SFGs using the global deblending technique and found
excellent agreement with the median results. The $24\,{\rm \mu m}$ flux densities
were measured on the stacked images using an aperture radius of 3.5$\arcsec$
with aperture corrections applied following the MIPS handbook. For the radio
fluxes we adopted the central pixel values. In both cases the errors were
estimated from the rms of the background in the stacked images.


\section{Results}
\subsection{Panchromatic UV-to-radio SEDs of QGs}
The stacked MIPS $24\,{\rm \mu m}$, \textit{Herschel}, and radio flux densities of
the $z$- and $M_{\star}$-bins of QGs are listed in
Table~\ref{table:stack}. None of the QG samples are significantly
detected (i.e., S/N $>3$) in any of the \textit{Herschel} stacks.
The most massive ($M_{\star}\geqslant 10^{10.6}\,M_{\odot}$) QGs are
detected at all redshifts out to $z=3$ in the $24\,{\rm \mu m}$ stacks (S/N $\sim5\mbox{--}26$)
and out to $z=1.5$ in the radio stacks (S/N $\sim4\mbox{--}10$). The intermediate-mass
QGs ($M_{\star} < 10^{10.6}\,M_{\odot}$) are detected at $24\,{\rm \mu m}$
(S/N $\sim5\mbox{--}20$) but not in the radio (S/N $\leqslant3$) in all relevant (i.e.,
mass-complete) redshift bins. As expected, $S_{24}$ and $S_{\rm radio}$ decrease
with $z$ (cosmic dimming) and increase with $M_{\star}$.

Figure \ref{fig:sed} summarises our constraints on the SEDs of QGs at mid-, far-IR, and radio wavelengths along with the median UV-to-near-IR SEDs. 
Note, the \textit{Herschel} non-detections are shown as
$3\sigma$ upper limits (i.e.\ $3\sigma_{\rm map}/\sqrt{N_{\rm stack}}$, where
$\sigma_{\rm map}$ is the map rms noise and $N_{\rm stack}$ the number of galaxies
in the stack).  For comparison we show the SED template of a dust-free
elliptical galaxy \citep{Bruzual2003} scaled to match the UV-to-near-IR median
photometry of the QGs (Figure \ref{fig:sed}, left).  The model,
which represents pure stellar emission, is insufficient to fully account
for the stacked $24\,{\rm \mu m}$ flux densities.
If we instead fit a modified black-body law (see details in Section \ref{sec:ms}) to the \textit{Herschel} limits, add radio emission
(a power-law with slope $\alpha=-0.8$) such that the radio-FIR correlation
\citep{Ivison2010b} is fulfilled, we still fall short of the stacked 1.4\,GHz fluxes (Figure \ref{fig:sed}, right panels).  The implications of
this excess emission at $24\,{\rm \mu m}$ and 1.4\,GHz are discussed in
Section~\ref{sec:agn}.

\capstartfalse
\begin{deluxetable*}{cccccc} 
	\tablecolumns{5}
	\tablecaption{\textit{Spitzer} $24\,{\rm \mu m}$ and \textit{Herschel} detected fractions for QGs}
	\tablehead{ 
	 & \multicolumn{4}{c}{log($M_{\star}/M_{\odot}$)} \\
	 &
	 11 -- 12.2 & 
	 10.6 -- 11 & 
	 10.2 -- 10.6 & 
	 9.8 -- 10.2  \\ \\
	 Redshift & $f_{\rm QG, 24}$ ($f_{\rm QG, H}$) & $f_{\rm QG, 24}$ ($f_{\rm QG, H}$) &
	 $f_{\rm QG, 24}$ ($f_{\rm QG, H}$) & $f_{\rm QG, 24}$ ($f_{\rm QG, H}$)
	}
	\startdata
	0.1 -- 0.5 & 0\% (0\%) & 0.4\% (0.2\%) & 0\% (0\%) & 0.2\% (0\%) \\ 
	0.5 -- 1.0 & 4.6\% (1.4\%) & 4.0\% (1.0\%) & 2.2\% (0.3\%) & 0.4\% (0.1\%) \\ 
	1.0 -- 1.5 & 9.9\% (2.4\%) & 5.3\% (1.0\%) & 2.4\% (0\%) & 0.6\% (0\%) \\ 
	1.5 -- 2.0 & 8.8\% (1.5\%) & 9.0\% (1.4\%)  & 7.5\% (1.0\%) & \nodata \\
	2.0 -- 2.5 & 19.4\% (6.0\%) & 17.4\% (2.5\%) & \nodata & \nodata \\
	2.5 -- 3.0 & 13.3\% (2.7\%) & \nodata & \nodata & \nodata 
	\enddata
	\label{table:detfrac}
	\tablecomments{
	$f_{\rm QG, 24}$ is the fraction of QGs (classified by their NUV$-$r and r$-$J
	colors) with $24\,{\rm \mu m}$-inferred SFRs $>20\,M_{\odot}$yr$^{-1}$.
	$f_{\rm QG, H}$ is the fraction of QGs fulfilling the above $24\,{\rm \mu m}$ criterion that are also detected in at least two \textit{Herschel} PACS+SPIRE bands (S/N $\geqslant 5$).
	}
\end{deluxetable*}
\capstarttrue


\subsection{Where do QGs lie relative to the SFGs on the SFR-$M_{\star}$ relation?} \label{sec:ms}
The mid-, far-IR and radio stacks each provide an independent measurement of the
dust-obscured SFRs in our QGs.  Firstly, we estimate the 8 to 1000\,$\mu$m
rest-frame IR luminosity ($L_{\rm IR}$) from $S_{24}$ using the calibration by
\citet{Rujopakarn2013}, including the 0.13\,dex scatter of the calibration in the
error budget.  Independent $L_{\rm IR}$ upper limits are then obtained by
redshifting and scaling a modified black-body model to the \textit{Herschel}
$3\sigma$ upper limits using the IDL code of \citet{Casey2012}.  We use an
optically thick, modified black-body law with a fixed dust temperature
($T_{\rm dust} = 30\,{\rm K}$) and emissivity 1.5.  Note that $L_{\rm IR}$ is insensitive to
$T_{\rm dust}$, as it only varies by less than a factor of two if we assume
$T_{\rm dust} = 15\,{\rm K}$ or 50\,K instead.  For each of our QG samples, $L_{\rm IR}$ is
estimated in the above manner using the median $z_{\rm phot}$ (listed in
Table~\ref{table:stack}), and subsequently converted into an obscured SFR using
the $L_{\rm IR}$-SFR calibration by \citet{Kennicutt1998} adjusted to the IMF used
in this work.  Assuming that all the radio emission originates from SF, and a
radio spectral index $\alpha =-0.8$, rest-frame 1.4\,GHz luminosities ($L_{\rm 1.4\,GHz}$) are derived from the radio stacks and subsequently converted to SFR$_{\rm radio}$ using
the $L_{\rm 1.4\,GHz}$-SFR calibration by \citet{Bell2003}.

The \textit{Herschel} $L_{\rm IR}$ upper limits and the (specific) SFRs for QGs as a
function of $M_{\star}$ and $z$ are listed in Table~\ref{table:stack}. The
\textit{Herschel} upper limits put stringent constraints on the dust-obscured
SFR: $<1\,M_{\odot}$yr$^{-1}$ at $z<1.5$ and at most $<18\,M_{\odot}$yr$^{-1}$,
i.e., sSFR $\leqslant 10^{-(10\mbox{--}12)}$yr$^{-1}$, 
across all $z$ and $M_{\star}$ bins.
These limits are consistent with the quiescence inferred from the unobscured SFR from
UV-to-IRAC SED fits ($0.01 \mbox{--} 1.2\,M_{\odot}$ yr$^{-1}$, 
see Table~\ref{table:stack}).  
QGs form stars at a very modest rate ($>$10$\times$
lower than SFGs, Figure~\ref{fig:ms}).  As a consistency check, we find that the
stacked \textit{Herschel} flux densities of SFGs obtained from global deblending and stacking \citep{Kurczynski2010} are in good agreement with those from median combination,
and we recover the SFR-$M_{\star}$ sequence found in a recent compilation of
similar measurements (\citealt{Speagle2014}, Figure~\ref{fig:ms}).


\subsection{Do QGs host AGN?}\label{sec:agn}
SFR$_{24}$ are consistent with SFR$_{\rm H}$, except at $z<1.5$ in which SFR$_{24}$
is higher than SFR$_{\rm H}$ (by as much as 5$\times$), as shown in
Figure~\ref{fig:ms}.  This discrepancy could be explained by the following
factors not related to recent SF: (1) The Rayleigh-Jeans tail of the stellar
photospheric emission, which is dominated by red giants; (2) Circumstellar dust
envelops of asymptotic giant branch (AGB) stars \citep{Knapp1992, Piovan2003};
(3) Interstellar (cirrus) dust heated by evolved stellar populations
\citep[e.g.,][]{Bendo2012}; (4) Polycyclic aromatic hydrocarbon emission associated with (2) and (3)
\citep{Kennicutt1998, Bendo2008}; (5) Warm dust heated by the obscured AGN
\citep{Daddi2007}.  The first four factors are viable for galaxies with
intermediate-old stellar populations ($>1$\,Gyr, \citealt{Salim2009}).  The
elliptical galaxy template from stellar population synthesis models
\citep{Bruzual2003}, which accounts for only (1) and to some extend (2), cannot
fully reproduce the observed $S_{24}$ at least for $z>0.5$
(Figure~\ref{fig:sed}, left).  This suggests that AGN and/or dust heating from
evolved stellar populations are likely responsible for the low levels of
$L_{\rm IR}$ of QGs \citep{Salim2009, Bendo2012, Fumagalli2013, Utomo2014}.  The
relative contribution of the above factors depends heavily on the evolution
models of AGB stars, dust grain models and interstellar radiation strength,
which are actively debated and beyond the scope of this Letter.  While we cannot
discern the relative contributions of dust heating from these factors using the
data in hand, we note that the SFR$_{24}$ are likely upper limits if non-SF
processes contribute significantly to $S_{24}$.

It is interesting that SFR$_{\rm radio}$ is systematically higher than SFR inferred
from $24\,{\rm \mu m}$ and \textit{Herschel} as well as UV-to-IRAC SED fits, up to two
orders of magnitude in the most extreme case.  Compared to the total (obscured +
unobscured) SFR inferred from other indicators (\textit{Herschel}, MIPS, UV-to-IRAC
SED fitting), the median $S_{\rm radio}$ are inconsistent with originating from SF
alone.  This is reflected in the low radio index ($q_{24} \equiv $ log($S_{24} /
S_{\rm radio}$) ) listed in Table~\ref{table:stack} compared to SFGs with typical
values of 1.5 -- 3 \citep[e.g.,][]{Ivison2010b}.  $L_{\rm 1.4\,GHz}$ [W Hz$^{-1}$]
increases with redshift from $10^{21.5}$ at $z\sim0.4$ to $10^{23.7}$ at
$z\sim2.6$ for the most massive QGs, where the radio excess is the most
prominent (see Figure~\ref{fig:sed} right panel insets).  Based on the FIR-radio correlation
presented in \citet{Ivison2010b} and including the radio-detected QGs in the
stack, we estimate that 20-90\% of $L_{\rm 1.4\,GHz}$ arises from non-SF processes.
This fraction is significantly higher for more massive QGs as shown in
Figure~\ref{fig:sed}, although we note that if we adopt a more conservative
\textit{Herschel} upper limit for the non-detection, the fraction will be lower.
Our results indicate that low-luminosity radio AGN may be widespread among
massive QGs, echoing the reciprocatory that massive QGs are the preferential
hosts for low-luminosity radio AGNs \citep[e.g.,][]{Smolcic2009}.  However, it
is not straightforward to use the median stacked radio luminosity to constrain
the heating rate of radio-AGN feedback, without prior assumption of the duty
cycle which is not well quantified.


\section{Discussion} \label{sec:discussion}
We reject the null hypothesis that the red colors of QGs are due to strong
obscured SF, based on a deep FIR stacking analysis. QGs have truly low SFRs and
evolved stellar populations, as expected from their low unobscured SFRs measured
from the UV continua. The average sSFRs of QGs are at least 1\,dex lower than those of SFGs out
to $z=3$.  The stacked $24\,{\rm \mu m}$ and radio emissions cannot be completely
accounted for by low levels of dust-obscured SFR nor stellar emissions,
suggesting that low-luminosity AGN may be present in QGs.

Comparing with \citet{Fumagalli2013}, who performed $24\,{\rm \mu m}$
stacking on 309 QGs with $M_{\star} \geqslant 10^{10.3}\,M_{\odot}$, our $S_{24}$ are slightly higher (5--26\,$\mu$Jy vs 2--3\,$\mu$Jy).
Their sample is drawn from a smaller survey area equivalent to 11\% of the UltraVISTA field,
and therefore the discrepancy is likely explained by the fact that their sample is dominated by lower mass galaxies, which have lower $S_{24}$.  Nevertheless, we arrive at similar
conclusions --- QGs do not host strong obscured SF, and dust heating by
evolved stellar populations may be significant at the low levels of $L_{\rm IR}$
observed.  Our results indicate that $z\gtrsim2$ QGs have average
$L_{\rm IR} \leqslant 10^{11.2}\,L_{\odot}$, i.e., $\geqslant0.8\,$dex below the
ULIRG threshold.  When we repeat our stacking analysis including QGs detected
at $24\,{\rm \mu m}$ following the definition of \citet{Viero2013}, we obtain higher
stacked mid- and far-IR emission, in broad agreement with their results.  As QGs have
higher $24\,{\rm \mu m}$ and \textit{Herschel} detection fractions at $z\gtrsim2$ (up to
19\% and 6\%, respectively, see Table~\ref{table:detfrac} and
Section~\ref{sec:data}), the inclusion of the quoted fractions of
$L_{\rm IR} > 10^{13}\,L_{\odot}$ sources boosts the stacked FIR emission of massive
QGs at $z\gtrsim2$ to be comparable to ULIRGs.

We reaffirm that a population of truly quiescent galaxies is already in place by
$z=3$.  This corroborates the need for powerful quenching mechanisms to terminate 
star formation in galaxies.  While environmental quenching may be dominant
for intermediate-mass QGs \citep{Peng2010}, stacking analyses at radio (this
work) and X-ray \citep{Olsen2013} wavelengths reveal that massive QGs harbor
low-luminosity AGN.  AGN provide a viable mechanism for quenching SF in
galaxies, as supported by the enhanced AGN fraction among transitory objects
between SFGs and QGs \citep[e.g.,][]{Barro2014a}.  After galaxies are quenched,
the AGN may then proceed to ``maintenance mode'' suppressing further SF through
a feedback cycle \citep{Schawinski2009a, Best2012}.  With upcoming surveys it will be possible to
conduct a complete census of AGN to sample the entire feedback duty cycle and
constrain their energetics, in order to quantify their role in quenching star
formation in galaxies.


\acknowledgments{}
We thank the COSMOS, UltraVISTA, PEP, and HerMES collaborations for providing the data used here.
Dark Cosmology Centre is funded by DNRF. 
AM thanks Anna Gallazzi, Mark Sargent, Ryan Quadri, Brian Lemaux, and Julie Wardlow for helpful discussions.
TRG acknowledges support from an STFC Advanced Fellowship.
Support for BM was provided by the DFG priority program 1573 ``The physics of the interstellar medium''.
AK acknowledges support by the Collaborative Research Council 956, sub-project A1, funded by the Deutsche Forschungsgemeinschaft (DFG).
CMC acknowledges support from a McCue Fellowship through the University of California, Irvine's Center for Cosmology.
VS is funded by the European Union's Seventh Frame-work program (grant agreement 337595).
Based on data products from observations made with ESO Telescopes at the La Silla Paranal Observatory under ESO programme ID 179.A-2005 and on data products produced by TERAPIX and the Cambridge Astronomy Survey Unit on behalf of the UltraVISTA consortium.

\bibliographystyle{apj}

\newpage
\clearpage
\pagestyle{empty}
\hoffset0.7in 
\voffset-0.6in
\capstartfalse
\begin{landscape}
\begin{deluxetable}{ccccccccccccccccc}
	\tablecolumns{17}
	\tablecaption{
	Stacked flux densities of \textit{Spitzer}/MIPS $24\,{\rm \mu m}$,
	\textit{Herschel} (PACS+SPIRE), and VLA 1.4\,GHz and the inferred SFRs for QGs
	}
	  \tablehead{ 
	  \colhead{Redshift} &
	  \colhead{$\overline{z_{\rm phot}}$} &
	  \colhead{$N_{\rm stack}$} &
	  \colhead{S$_{24\,{\rm \mu m}}$} &
 	  \colhead{S$_{100\,{\rm \mu m}}$} &
	  \colhead{S$_{160\,{\rm \mu m}}$} &
 	  \colhead{S$_{250\,{\rm \mu m}}$} &
  	  \colhead{S$_{350\,{\rm \mu m}}$} &
	  \colhead{S$_{500\,{\rm \mu m}}$} &
	  \colhead{S$_{\rm radio}$} &
	  \colhead{SFR$_{\rm SED}$} &
	  \colhead{SFR$_{24}$} &
	  \colhead{SFR$_{\rm H}$} &
	  \colhead{SFR$_{\rm radio}$} &
	  \colhead{log($L_{\rm IR, H}$)} &
	  \colhead{log(sSFR$_{\rm H})$} &
	  \colhead{$q_{24}$} \\
	  \colhead{} & \colhead{} & \colhead{} &
	  \colhead{[$\mu$Jy]} & \colhead{[mJy]} & \colhead{[mJy]} &
	  \colhead{[mJy]} & \colhead{[mJy]} & \colhead{[mJy]} & \colhead{[$\mu$Jy]} &
	  \colhead{[$M_{\odot}$yr$^{-1}$]} & \colhead{[$M_{\odot}$yr$^{-1}$]} & 
	  \colhead{[$M_{\odot}$yr$^{-1}$]} & \colhead{[$M_{\odot}$yr$^{-1}$]} &
	  \colhead{log[$L_{\odot}$]} & \colhead{log[yr$^{-1}$]} & \colhead{}
	  }
	\startdata	

\cutinhead{log($M_{\star}/M_{\odot})$ = 11 -- 12.2 (median = 11.1)}
0.1 -- 0.5 & 0.4 & 229 & 32.3$\pm$1.5 & 0.4$\pm$0.3 & 0.3$\pm$0.7 & 0.4$\pm$0.5 & 0.0$\pm$0.7 & -0.1$\pm$0.9 & 5.2$\pm$1.3 & 0.03 & 0.4$\pm$0.2 & $<$0.2 & 1.2$\pm$0.2 & $<$9.2&$<-11.9$ & 0.8 \\ 
0.5 -- 1.0 & 0.8 & 1222 & 21.3$\pm$0.8 & 0.0$\pm$0.1 & 0.2$\pm$0.3 & -0.2$\pm$0.2 & -0.2$\pm$0.3 & -0.1$\pm$0.4 & 5.2$\pm$0.5 & 0.13 & 1.6$\pm$0.5 & $<$0.3 & 6.2$\pm$0.6 & $<$9.5&$<-11.7$ & 0.6 \\ 
1.0 -- 1.5 & 1.2 & 733 & 15.9$\pm$0.9 & 0.1$\pm$0.2 & -0.2$\pm$0.4 & -0.2$\pm$0.3 & 0.2$\pm$0.4 & 0.4$\pm$0.5 & 3.6$\pm$0.6 & 0.42 & 3.3$\pm$1.1 & $<$0.8 & 11.1$\pm$2.0 & $<$9.9&$<-11.2$ & 0.6 \\ 
1.5 -- 2.0 & 1.7 & 288 & 8.8$\pm$1.1 & -0.4$\pm$0.3 & -0.4$\pm$0.<6 & -0.8$\pm$0.5 & -0.7$\pm$0.6 & -0.1$\pm$0.8 & 2.3$\pm$0.9 & 0.35 & 3.0$\pm$1.1 & $<$2.9 & 16.3$\pm$6.6 & $<$10.4&$<-10.7$ & 0.6 \\ 
2.0 -- 2.5 & 2.2 & 174 & 13.4$\pm$1.5 & -0.4$\pm$0.4 & 0.5$\pm$0.8 & -0.4$\pm$0.6 & -0.1$\pm$0.8 & 0.3$\pm$1.0 & 4.2$\pm$1.2 & 0.71 & 6.9$\pm$2.5 & $<$7.1 & 59.6$\pm$17.4 & $<$10.8&$<-10.3$ & 0.5 \\ 
2.5 -- 3.0 & 2.6 & 65 & 10.8$\pm$2.2 & -0.4$\pm$0.6 & 0.1$\pm$1.3 & -0.6$\pm$1.0 & -0.1$\pm$1.4 & -0.0$\pm$1.6 & 6.5$\pm$2.2 & 1.20 & 11.4$\pm$4.8 & $<$17.6 & 145.0$\pm$49.2 & $<$11.2&$<-9.9$ & 0.2 \\ 

\cutinhead{log($M_{\star}/M_{\odot}$) = 10.6 -- 11.0 (median = 10.8)}
0.1 -- 0.5 & 0.4 & 502 & 25.2$\pm$1.0 & 0.6$\pm$0.2 & 1.0$\pm$0.5 & 0.0$\pm$0.4 & -0.7$\pm$0.5 & -0.6$\pm$0.6 & 3.5$\pm$0.8 & 0.02 & 0.3$\pm$0.1 & $<$0.1 & 0.8$\pm$0.1 & $<$9.0&$<-11.8$ & 0.9 \\ 
0.5 -- 1.0 & 0.8 & 2167 & 14.2$\pm$0.6 & 0.0$\pm$0.1 & -0.0$\pm$0.2 & -0.7$\pm$0.2 & -0.6$\pm$0.2 & -0.2$\pm$0.3 & 1.7$\pm$0.4 & 0.10 & 1.0$\pm$0.3 & $<$0.2 & 2.0$\pm$0.3 & $<$9.3&$<-11.4$ & 0.9 \\ 
1.0 -- 1.5 & 1.2 & 1646 & 8.4$\pm$0.5 & -0.1$\pm$0.1 & -0.4$\pm$0.3 & -1.1$\pm$0.2 & -1.1$\pm$0.3 & -0.8$\pm$0.3 & 1.6$\pm$0.4 & 0.23 & 1.5$\pm$0.5 & $<$0.5 & 4.8$\pm$1.2 & $<$9.7&$<-11.0$ & 0.7 \\ 
1.5 -- 2.0 & 1.7 & 516 & 6.0$\pm$0.8 & -0.2$\pm$0.2 & -0.3$\pm$0.5 & -1.5$\pm$0.3 & -1.5$\pm$0.5 & -0.9$\pm$0.6 & 2.1$\pm$0.7 & 0.30 & 1.9$\pm$0.7 & $<$2.2 & 15.3$\pm$5.1 & $<$10.3&$<-10.4$ & 0.5 \\ 
2.0 -- 2.5 & 2.2 & 295 & 8.5$\pm$1.2 & -0.2$\pm$0.3 & 0.2$\pm$0.6 & -0.9$\pm$0.5 & -0.6$\pm$0.6 & -0.4$\pm$0.8 & 2.3$\pm$1.0 & 0.65 & 4.6$\pm$1.7 & $<$5.9 & 36.0$\pm$14.8 & $<$10.8&$<-10.0$ & 0.6 \\ 

\cutinhead{log($M_{\star}/M_{\odot}$) = 10.2 -- 10.6 (median = 10.4)}
0.1 -- 0.5 & 0.4 & 699 & 12.8$\pm$0.8 & 0.5$\pm$0.2 & 1.0$\pm$0.4 & -0.9$\pm$0.3 & -1.5$\pm$0.4 & -1.2$\pm$0.5 & 2.6$\pm$0.7 & 0.01 & 0.1$\pm$0.1 & $<$0.1 & 0.6$\pm$0.1 & $<$8.9&$<-11.4$ & 0.7 \\ 
0.5 -- 1.0 & 0.8 & 2281 & 10.1$\pm$0.5 & 0.1$\pm$0.1 & -0.1$\pm$0.2 & -1.0$\pm$0.2 & -1.2$\pm$0.2 & -0.8$\pm$0.3 & 0.9$\pm$0.3 & 0.06 & 0.7$\pm$0.2 & $<$0.2 & 1.2$\pm$0.3 & $<$9.3&$<-11.1$ & 1.1 \\ 
1.0 -- 1.5 & 1.2 & 1199 & 5.4$\pm$0.6 & -0.1$\pm$0.1 & 0.0$\pm$0.3 & -1.4$\pm$0.2 & -1.2$\pm$0.3 & -0.7$\pm$0.4 & 1.3$\pm$0.5 & 0.15 & 0.9$\pm$0.3 & $<$0.6 & 4.0$\pm$1.4 & $<$9.8&$<-10.6$ & 0.6 \\ 
1.5 -- 2.0 & 1.7 & 432 & 5.4$\pm$1.0 & -0.2$\pm$0.2 & 0.1$\pm$0.5 & -1.4$\pm$0.4 & -1.3$\pm$0.5 & -0.8$\pm$0.6 & 2.2$\pm$0.8 & 0.31 & 1.7$\pm$0.7 & $<$2.4 & 16.2$\pm$5.6 & $<$10.4&$<-10.0$ & 0.4 \\ 

\cutinhead{log($M_{\star}/M_{\odot}$) = 9.8 -- 10.2 (median = 10.0)}
0.1 -- 0.5 & 0.4 & 583 & 13.1$\pm$0.8 & 0.3$\pm$0.2 & 0.1$\pm$0.4 & -1.8$\pm$0.3 & -2.1$\pm$0.5 & -1.9$\pm$0.5 & 1.8$\pm$0.7 & 0.01 & 0.1$\pm$0.1 & $<$0.1 & 0.5$\pm$0.1 & $<$8.9&$<-11.0$ & 0.9 \\ 
0.5 -- 1.0 & 0.8 & 1303 & 6.4$\pm$0.6 & -0.1$\pm$0.1 & -0.1$\pm$0.3 & -1.2$\pm$0.2 & -1.3$\pm$0.3 & -0.7$\pm$0.4 & 0.8$\pm$0.4 & 0.06 & 0.4$\pm$0.1 & $<$0.3 & 1.1$\pm$0.4 & $<$9.4&$<-10.6$ & 0.9 \\ 
1.0 -- 1.5 & 1.2 & 673 & 4.5$\pm$0.7 & 0.1$\pm$0.2 & -0.4$\pm$0.4 & -1.3$\pm$0.3 & -1.3$\pm$0.4 & -0.7$\pm$0.5 & 1.2$\pm$0.6 & 0.21 & 0.7$\pm$0.3 & $<$0.8 & 3.6$\pm$1.8 & $<$9.9&$<-10.1$ & 0.6 \\ 

\enddata
	\tablecomments{
	The median redshifts ($\overline{z_{\rm phot}}$), the number of QGs stacked ($N_{\rm stack}$), and the stacked flux densities are listed.
	SFR$_{\rm SED}$ is the median SFR from the UV-to-IRAC SED fitting.
	We infer SFRs from the stacked flux densities (Section~\ref{sec:ms}),
	assuming that the $24\,{\rm \mu m}$ and radio emissions originate from SF only.
	The IR luminosity and specific SFR upper limits ($L_{\rm IR,H}$ and sSFR$_{\rm H}$) inferred from the \textit{Herschel} upper limits are shown in logarithmic units.
	The radio index $q_{24}$ is computed as log($S_{24\,{\rm \mu m}}/S_{\rm radio}$).
	}
	\label{table:stack}
\end{deluxetable}
\clearpage
\end{landscape}
\capstarttrue

\end{document}